\documentclass[aps, reprint,
	10pt, notitlepage,
    floats, floatfix,
	amsmath, amssymb, amsfonts, 
	superscriptaddress,
	showpacs, showkeys,
	nofootinbib,prd
]{revtex4-2}
\usepackage{aasmacros}
\usepackage{graphicx} 
\usepackage[usenames,dvipsnames]{xcolor}
\usepackage{xspace} 
\usepackage{bm} 
\usepackage[utf8]{inputenc} 
\xdefinecolor{mylinkcolor}{rgb}{0,0,0.5}
\usepackage[
	bookmarksnumbered, bookmarksopen, bookmarksopenlevel=2,
	breaklinks=true,
	colorlinks=true, filecolor=mylinkcolor, citecolor=mylinkcolor,
	linkcolor=mylinkcolor, urlcolor=mylinkcolor, menucolor=mylinkcolor,
]{hyperref}
\usepackage{verbatim}
\usepackage{color}
\usepackage[dvipsnames]{xcolor}
\usepackage[normalem]{ulem}

\newcommand{\Ins}{\affiliation{Dipartimento di Scienza e Alta Tecnologia, Università dell’Insubria, via Valleggio 11, I-22100 Como, Italy}}
\newcommand{\Bic}{\affiliation{Dipartimento di Fisica “G. Occhialini”, Università degli Studi di Milano-Bicocca, Piazza della Scienza 3, 20126 Milano, Italy}}
\newcommand{\Infn}{\affiliation{INFN, Sezione di Milano-Bicocca, Piazza della Scienza 3, 20126 Milano, Italy}}
\newcommand{\bham}{\affiliation{Institute for Gravitational Wave Astronomy \& School of Physics and
Astronomy, University of Birmingham, Birmingham, B15 2TT, UK}}
\newcommand{\inaf}{\affiliation{INAF – Osservatorio Astronomico di Brera, via E. Bianchi
46, 23807, Merate, Italy}}

\begin{document}

\title{A weakly-parametric approach to stochastic background inference in LISA}

\author{Federico~Pozzoli\orcidlink{0009-0009-6265-584X}} \Ins
\email{fpozzoli@uninsubria.it}
\author{Riccardo~Buscicchio\orcidlink{0000-0002-7387-6754}}\Bic \bham \Infn \inaf 
\author{Christopher~J.~Moore\orcidlink{0000-0002-2527-0213}}\bham
\author{Francesco Haardt\orcidlink{0000-0003-3291-3704}}\Ins \Infn \inaf
\author{Alberto Sesana\orcidlink{0000-0003-4961-1606}}\Bic \Infn \inaf

\newcommand{\FP}[1]{{\color{Red} {[\textbf{FP}}: #1]}}
\newcommand{\RB}[1]{{\color{teal} {[\textbf{RB}}: #1]}}
\newcommand{\cm}[2]{{\color{blue} {\textbf{CM}}:}{\color{red}{\sout{#1}}}{\color{green}{#2}}}

\newcommand\orcidlink[1]{\href{https://orcid.org/#1}{\includegraphics[scale=0.006]{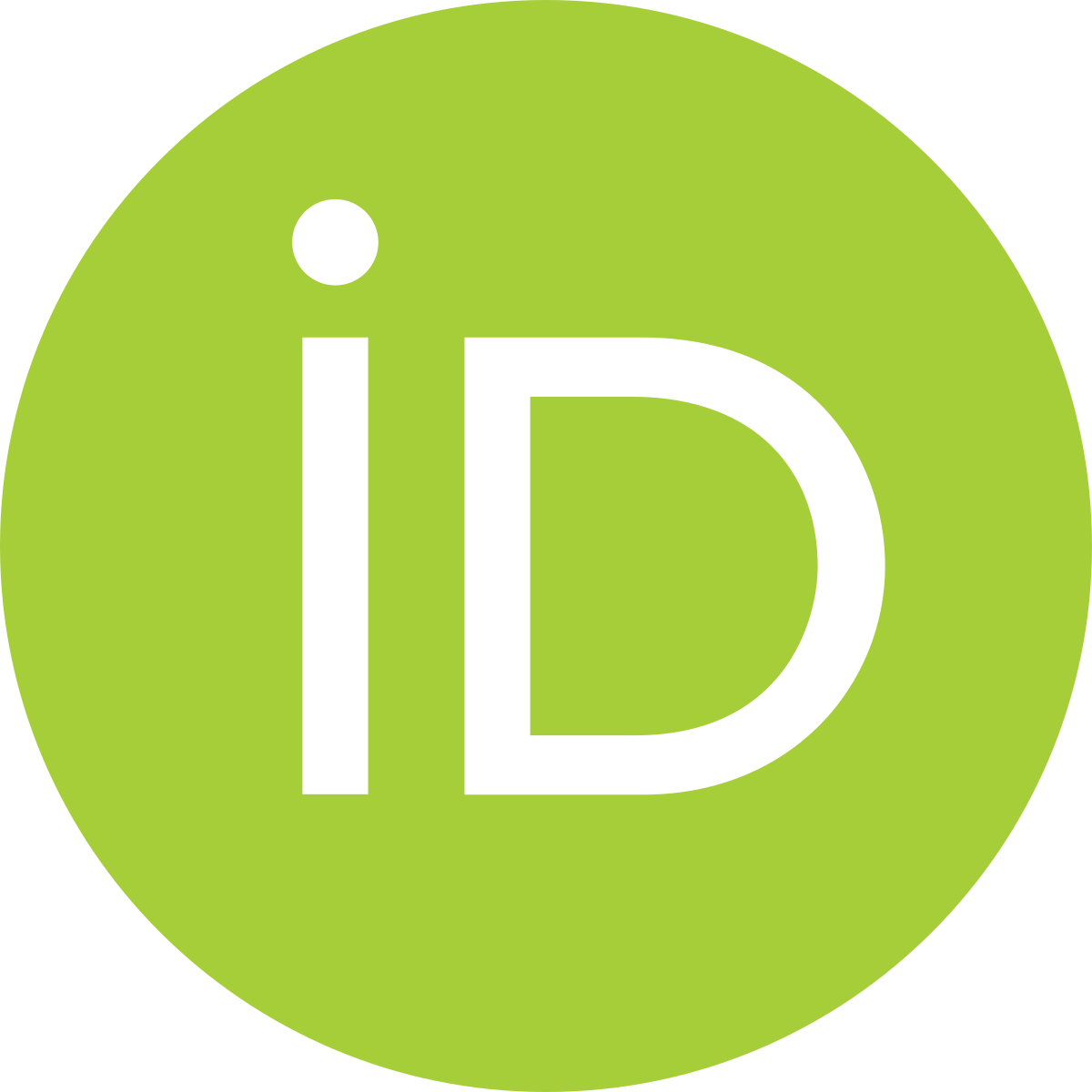}}}
\begin{abstract}
Detecting stochastic gravitational wave backgrounds (SGWBs) with The Laser Interferometer Space Antenna (LISA) is among the mission science objectives.
Disentangling SGWBs of astrophysical and cosmological origin is a challenging task, further complicated by the noise level uncertainties.
In this study, we introduce a Bayesian methodology to infer upon SGWBs, taking inspiration from Gaussian stochastic processes.
We investigate the suitability of the approach for signal of unknown spectral shape. We do by discretely exploring the model hyperparameters,  a first step towards a more efficient transdimensional exploration. 
We apply the proposed method to a representative astrophysical scenario: the inference on the astrophysical foreground of Extreme Mass Ratio Inspirals, recently estimated in~\cite{Pozzoli2023}.
We find the algorithm to be capable of recovering the injected signal even with large priors, while simultaneously providing estimate of the noise level.
\end{abstract}

\maketitle

\section{Introduction}
\label{sec:intro}
The Laser Interferometer Space Antenna (LISA)~\cite{LISA2017} is a groundbreaking mission for the detection of Gravitational Waves (GW) from space. 
Throughout the nominal $4$ years of duration of operations, an abundance of individually resolvable sources will be available.
A primary focus of LISA will be the detection of several tens of massive black-hole binary (MBHBs) mergers \cite{Klein2016,Sesana2004}, spanning masses from approximately $10^4$ to $10^7$ times that of the Sun. These mergers are anticipated to have high Signal-to-Noise Ratios (SNRs), reaching up to several thousands. 
Additionally, LISA will identify several to a few hundred extreme mass ratio inspirals (EMRIs) annually~\cite{Babak2017}, along with tens of thousands of Galactic white dwarf binaries \cite{Korol2020}.

Besides individual sources, the Universe is pervaded by a stochastic gravitational wave background (SGWB), which arises from an incoherent superposition of GWs originating from numerous unresolved or weak sources.
At nano-Hz frequencies, the first evidence for an unresolved GW signal of astrophysical and/or cosmological origin \cite{2023arXiv230616227A,2023ApJ...951L..11A} has been recently been reported by the European pulsar timing array \cite{2023A&A...678A..50E}, NANOGrav \cite{nanograv}, the Parkes Pulsar timing array \cite{2023ApJ...951L...6R} and the Chinese pulsar timing array \cite{2023RAA....23g5024X}. 
The SGWB in LISA may come from both cosmological and astrophysical sources. Cosmologically~\cite{Caprini2015}, these waves could stem from primordial quantum fluctuations~\cite{Ricciardone2017}, potentially amplified during cosmic inflation. Furthermore, processes such as first-order phase transitions~\cite{Gowling2021,Boileau2023}, interacting cosmic strings~\cite{Auclair2020}, and primordial black holes~\cite{Clesse2017} may also contribute to stochastic GW emissions. Astrophysically, the Galactic foreground~\cite{Nelemans2009} from double white dwarf binaries (DWDs) is expected to dominate, particularly in the $0.5-3$ mHz range. Additionally, both stellar-origin black-hole binaries (SOBHBs)~\cite{Babak2023} and EMRIs have the potential to generate non-negligible backgrounds~\cite{Pozzoli2023,Bonetti2020}.

The persistence of the SGWB in the collected data ties inextricably to the estimation of instrumental noise. Various efforts have been made to construct a null channel in LISA, that is a datastream effectively insensitive to GWs: however, in realistic instrumental setups, this won't be possible across the whole sensitivity band~\cite{Hartwig2023}.
 
Disentangling different SGWB components and instrumental noise provides a challenging data-analysis task.
Many approaches have been already proposed using approaches based on stochastic-sampling algorithm such as Markov Chain Monte Carlo (MCMC)\cite{mcmc} or nested sampling \cite{ns}. 
In particular, noise and signal reconstruction leveraging templated inference for both~\cite{Adams2014,Boileau2023}, or just one~\cite{Baghi2023,Caprini2019,Flauger2021} have been explored. 
In this study, we introduce a weakly-parametric approach by implementing an inference based on stochastic Gaussian processes, which can be applied in various contexts with great flexibility in the SGWB spectral shape. 

The paper is organized as follows. In Section~\ref{sec:astro} we provide a brief overview of the putative astrophysical SGWBs potentially observable with LISA. 
In Section~\ref{sec:lisa-data} we describe the underlying assumptions of our analysis. 
In Section~\ref{sec:stat-model} a detailed exposition of the core statistical model is presented. 
In Section~\ref{sec:result}, we report our parameter estimation results on a set of simulated SGWB signals. 
Finally, in Section~\ref{sec:conclusion}, we summarize our findings and outline the prospects for future developments towards a highly flexible inference on SGWBs. 
Unless otherwise specified, summation over repeated indices is assumed throughout.

\section{Astrophysical Backgrounds}
\label{sec:astro}
Astrophysical SGWBs are expected to form from the superposition of GWs from compact binary systems. 
However, other putative astrophysical SGWBs are expected to arise, e.g. from asymmetric supernova explosions~\cite{Rosca2023} or rapidly rotating neutron stars~\cite{Rosado2012}. 

Signals typically manifest as either isolated and faint,  falling below the detection threshold, or as overlapping ones. They pile up incoherently, hence resulting in a ``confused'' timeseries, whose morphology is too complex to tell individual signals apart. 
Moreover, the different astrophysical (and cosmological) SGWBs will overlap in the LISA data making it difficult to study them, individually~\cite{2021arXiv210506793B}. 
However, the different spectral shapes of each background component and the different sky distributions of Galactic (localised near the plane/buldge of the Milky Way) versus extragalactic (more isotropic) backgrounds can be used to separate them.
The dominant contributors to astrophysical SGWBs in the LISA sensitivity band are expected to originate mainly from three populations.
We summarize briefly their properties below. 

~~~\emph{Galactic DWDs.}~~~
The white dwarf binary Galactic population is expected to be formed by around $\tilde\times10^7$ systems~\cite{Timpano2006}. A significant fraction (up to $\sim10\%$) contributes with GW signals in the LISA band, i.e. at GW frequencies above $0.1 \rm mHz$~\cite{Korol2020}.
Thousands of such highly monochromatic systems will be resolved individually, some of which are already observed through their electromagnetic counterparts; these are known as verification binaries~\cite{Stroeer2006,Finch2023,Kupfer2023}.
Ongoing and future missions such as GAIA~\cite{Gaia} and the Vera Rubin Observatory~\cite{Vera} have the potential to unveil up to a few thousand more DWD systems.
The largest fraction of the DWDs will remain unresolved --only $\sim 0.1\%$ of DWDs are going to be resolved individually~\cite{Georgousi2023}-- and form a stochastic signal, commonly referred to as \emph{confusion noise}. 
This is expected to be larger than LISA instrumental noise the frequency range of $0.5-3$ mHz. 
The bulk of such foreground arises from systems located toward the Galactic center, hence the confusion noise is expected to be highly anisotropic.
Leveraging the SGWB sky distribution and the induced signal modulation due to LISA motion, is an opportunity to characterize the Galactic morphology~\cite{Breivik2020}, i.e. its disk, bulge, halo, and streams.

~~~\emph{SOBHBs.}~~~ 
SOBHBs will be detected by LISA during their early inspiral phase~\cite{Sesana2020}. A portion of these binaries will enter the final years of their inspiral and eventually transition into the frequency range detectable by LIGO, Virgo, and Kagra (LVK), thereby allowing for multiband gravitational-wave astronomy~\cite{Sesana2016,2022arXiv220403423K}.
Based on current LVK constraints, it is anticipated that at least a few of these binaries will be individually detected and characterized~\cite{2021PhRvD.104d4065B, 2022PhRvD.106j4034T, 2023PhRvD.108b3022D}, while the majority will contribute to the formation of a SGWB. The background is expected to be isotropic and power-law shaped with spectral index $2/3$ for the energy density power spectrum (or equivalently, $-7/3$ for the PSD), for quasi-circular binaries at leading order post-Newtonian approximation~\cite{Phinney2001}. 
Recently, accurate detectability of the SOBBHs background has been demostrated~\cite{Babak2023}, accounting for instrumental noise uncertainties and coexistence of DWDs foreground.

~~~\emph{EMRIs.}~~~
A stellar mass object orbiting around a MBH constitute an EMRI. These sources have not been observed through either electromagnetic or GW radiation, yet.
Nonetheless, GWs emitted by EMRIs are a primary focus for LISA, due to their potential association with Fast Radio Bursts in the electromagnetic spectrum~\cite{Li2023}.
Because of their very small mass ratios $\sim 10^{-6}-10^{-4}$, EMRIs evolve slowly: a considerable number of these systems will persist in the LISA band throughout the entire mission.
A large number of orbital cycles $\sim 10^4-10^5$ provides the opportunity to probe, in the test-particle limit, highly curved spacetime close to supermassive black holes (SMBHs).
More recently, EMRIs have been proposed as potential targets for strong gravitational lensing~\cite{Toscani2023}.
Most of the population will remain undetected, hence a SGWB is expected to form.
Unlike the SOBHBs population, the EMRI background spectral shape prediction is uncertain due to (i) the spread of each gravitational-wave signal over multiple harmonics of the orbital frequency and (ii) the uncertainty related to the populations. A systematic characterization of EMRI backgrounds, from astrophysically motivated model populations developed in~\cite{Babak2017} was recently presented in~\cite{Pozzoli2023}: the majority of models yield a bright SGWB with SNRs ranging between tens and thousands, contributing mostly to higher frequencies than the DWD background, namely $1-10$ mHz.

\section{Stochastic signals in LISA}
\label{sec:lisa-data}

The LISA constellation is composed of three spacecrafts arranged in a nearly equilateral configuration, respectively separated by $2.5 \times 10^6 \rm km$. Spacecrafts will trail the Earth motion by an angle of about $20^{\circ}$, lying on a plane inclined by $60$ degrees with respect to the orbital plane. Distances between spacecrafts will be modulated by incident GWs, inducing variations at the picometer level. 
Such variations will be detected by monitoring the frequency (or phase, equivalently) of laser beams exchanged between spacecrafts and comparing them to local reference lasers.
The primary source of noise will come from laser frequency fluctuations, many orders of magnitude higher than the target sensitivity. To suppress it, six raw Doppler measurements are suitably delayed and linearly combined in a set of time-delay interferometric (TDI) variables~\cite{Tinto2005}.
Various TDI scheme exist in literature, associated to different approximations of the LISA constellation orbits;
in this work we use TDI 1.5 combinations~\cite{Tinto2004}, assuming constant and unequal-length LISA arms.
A further linear combination of such TDI variables is used to define approximately noise-orthogonal datastreams.

We perform inference on such data, $\Tilde{d} = (\Tilde{A},\Tilde{E},\Tilde{T})$, the Fourier transforms of noise-orthogonal TDI channels.
We model the datastreams as the superposition of two independent stochastic processes: the SGWB, the instrumental noise. 
In this work we do not consider the contribution from the non-stationary Galactic foreground, whose addition we leave for future study.
At each frequency $f$ the correlation between data reads
\begin{equation}
    \left\langle\Tilde{d}_\alpha(f)\Tilde{d}^*_\beta(f)\right\rangle = S_{h,\alpha\beta}(f) + S_{n,\alpha\beta}(f),
    \label{eq1}
\end{equation}
where $\alpha$ and $\beta$ denote each TDI channel.
We assume throughout isotropic, Gaussian, stationary SGWBs with equal-weighted and uncorrelated polarizations.

The first component in the previous equation reads:
\begin{equation}
    S_{h,\alpha\beta}(f) = R_{\alpha\beta}(f)S_h(f).
    \label{eq:signal-spectrum}
\end{equation}
$S_h(f)$ represents the one-sided SGWB strain power spectral density (PSD)~\cite{Moore2015}, defined as
\begin{equation}
\label{eq:strainpsd}
    \left\langle h_{p}(f;\Omega)h^*_{p'}(f',\Omega')\right\rangle\!=\frac{1}{2}S_h(f)\delta(f-f')\delta(\Omega,\Omega')\delta_{pp'}
\end{equation} 
and $R_{\alpha\beta}$ is a $3\times3$ matrix obtained by combining $M_{\rm TDI}(f)$ ($3 \times 6$) and the single-link response matrix $G(f,t_0)$ ($6 \times 6$) as follows:
\begin{equation} \label{eq:single_link_response_matrix}
    R_{\alpha\beta}(f) = M_{\alpha i,\rm TDI}(f)G_{ij}(f,t_0)M^*_{j \beta, \rm TDI}(f).
\end{equation}
For completeness, we provide full derivation of $G(f,t_0)$ in appendix~\ref{Appendix}. 

The spectrum $S_{n,\alpha\beta}$ represents the single-sided noise power spectra in a single laser link.
Following~\cite{Baghi2023}, we further assume no correlations between individual links and the same PSD in each of them. Hence, PSDs in each channel depend uniquely on the transfer matrix $M_{\rm TDI}$ and on a common spectral model $S_n(f)$.
\begin{equation}
    S_{n,\alpha\beta}(f) = \frac{1}{2}S_n(f)M_{\alpha i, \rm TDI}(f)M_{i \beta,\rm TDI}^*(f)
    \label{eq:noise-spectrum}
\end{equation}
Specifically,  we include in $S_n$ secondary noise sources, namely the test mass (TM) and the optical metrology system (OMS) noise. The overall PSD is $S_n=2S_n^{\rm TM} + S_n^{\rm OMS}$ \cite{Hartwig2023}, where
\begin{align}
        S_{n}^{\rm TM}\left(f\right) &= A^2\left[1+\left(\frac{0.4 \rm mHz}{f} \right)^2\right] \left[1+\left(\frac{f}{8 \rm mHz} \right)^4\right] \nonumber\\
        &\times \left( \frac{1}{2\pi fc}\right)^2 \left( 10^{-30} \rm m^2/ \rm s^3\right),
\end{align}
and
\begin{equation}
    S_{n}^{\rm OMS}\!\left(f\right)\! =\!P^2\!\left[1\!+\!\left(\frac{2 \rm mHz}{f} \right)^4\right]\!\!
    \left( \frac{2\pi f}{c}\right)^2\!\!\!\!\left(10^{-24}\rm m^2 \!\cdot \rm s\right)
\end{equation}
with $A = 3$ and $P = 15$ \cite{LISA2018}.

Through TDI channels signal and noise PSDs, the SNR for an SGWB is readily defined as
\begin{equation}
    \mathrm{SNR}^2 = T\sum_{\alpha} \int^\infty_0 df\frac{S^2_{h,\alpha\alpha}}{S^2_{n,\alpha\alpha}}
\end{equation}
with $T$ the LISA observation time~\cite{2019PhRvD.100j4055S}.

\section{Inference model}\label{sec:stat-model}

In this study we only focus on the parameter estimation of SGWB and noise, thus we assume that all non-stochastic GW sources have been perfectly subtracted from the data.  
We choose to model both spectra -- introduced in Eqs.~\eqref{eq:signal-spectrum} and~\eqref{eq:noise-spectrum} -- with a weakly-parametric model, which allows us to explore a large family of spectral shapes. 
We do so by using expectation values of Gaussian processes.
A Gaussian process (GP) is a stochastic model, formally describing distributions over functions $g(x)$. It is parameterized by a mean function $m(x)$ and positive definite covariance function $k(x,x')$. Realizations from such process are denoted as follows
\begin{equation}
    g(x) \sim \mathcal{GP}(m(x),k(x,x')),
\end{equation}
The defining properties of GPs is that for any finite subset $X = \left\{x_1 \dots x_n\right\}$ of the domain for $x$, the marginal distribution is a multivariate Gaussian 
\begin{equation}
    \boldsymbol{g}(X) \sim \mathcal{N}(\boldsymbol{\mu},\boldsymbol{\Sigma}),
\end{equation}
with mean vector and covariance matrix defined by $\mu= m(x_i)$ and $\Sigma_{ij}= k(x_i,x_j)$, respectively.
Typically, GPs are used to flexibly incorporate some observed data in the distribution over $\boldsymbol{g}(X)$ and make predictions on new domain points $X_{*}$ as $\boldsymbol{g}(X_*)$. 
The joint probability distribution $p(\boldsymbol{g}({X}_*),\boldsymbol{g}({X}))$ is a multivariate normal distribution. In order to employ this formalism as a regression model, one needs the conditional probability $p(\boldsymbol{g}(X_*)| \boldsymbol{g}({X}))$. 
Upon suitable marginalization, the conditional distribution is also a multivariate normal distribution
\begin{equation}
    p(\boldsymbol{g}({X}_*)|\boldsymbol{g}({X})) = \mathcal{N}(\boldsymbol{\mu}\left({{X}_*|{X}}\right),\boldsymbol{\Sigma}\left({{X}_*|{X}}\right)),
    \label{eq:conditional-mvn}
\end{equation}
where
\begin{align} 
    \boldsymbol{\mu}\left({X_*|X}\right) 
    &= \boldsymbol{\mu}(X_*) \,+ \nonumber\\
    &+ \boldsymbol{\Sigma}\left({X_*|X}\right)\boldsymbol{\Sigma}\left({X,X}\right)^{-1}(\boldsymbol{g}({X})\! -\! \boldsymbol{\mu}({X})) \label{eq6}
\end{align}
and
\begin{align}
    \boldsymbol{\Sigma}\left({X_*|X}\right) &= \boldsymbol{\Sigma}\left({X_*,X_*}\right) + \nonumber \\
    &+\boldsymbol{\Sigma}\left({X_*,X}\right)\boldsymbol{\Sigma}^{-1}\left({X,X}\right)\boldsymbol{\Sigma}\left({X,X_*}\right),\label{eq:varGP}
\end{align}

are specified by the single kernel function $k(x,x')$. 
The matrix $\boldsymbol{\Sigma}$ is also referred to as \emph{kernel matrix}, and it models the covariance between each pair of its two arguments through the definition of the bivariate function $k$.
In order to be a valid covariance for the multivariate Gaussian in Eq.~\eqref{eq:conditional-mvn}, the kernel matrix must be symmetric and positive definite. A variety of kernels are available in literature, to capture different processes peculiarities. 
In this work, we consider the radial basis kernel function (RBF) defined by
\begin{equation}
\Sigma^{\rm RBF}_{ij}\left({X,Y}\right) = k_{\rm RBF}(x_i,y_j) = \exp{\left(-\frac{|x_i-y_j|^2}{2\sigma^2}\right)},
\end{equation}
where $\sigma$ is a model positive hyperparameter.

\begin{figure}
    \centering
    \includegraphics[width=0.95\columnwidth]{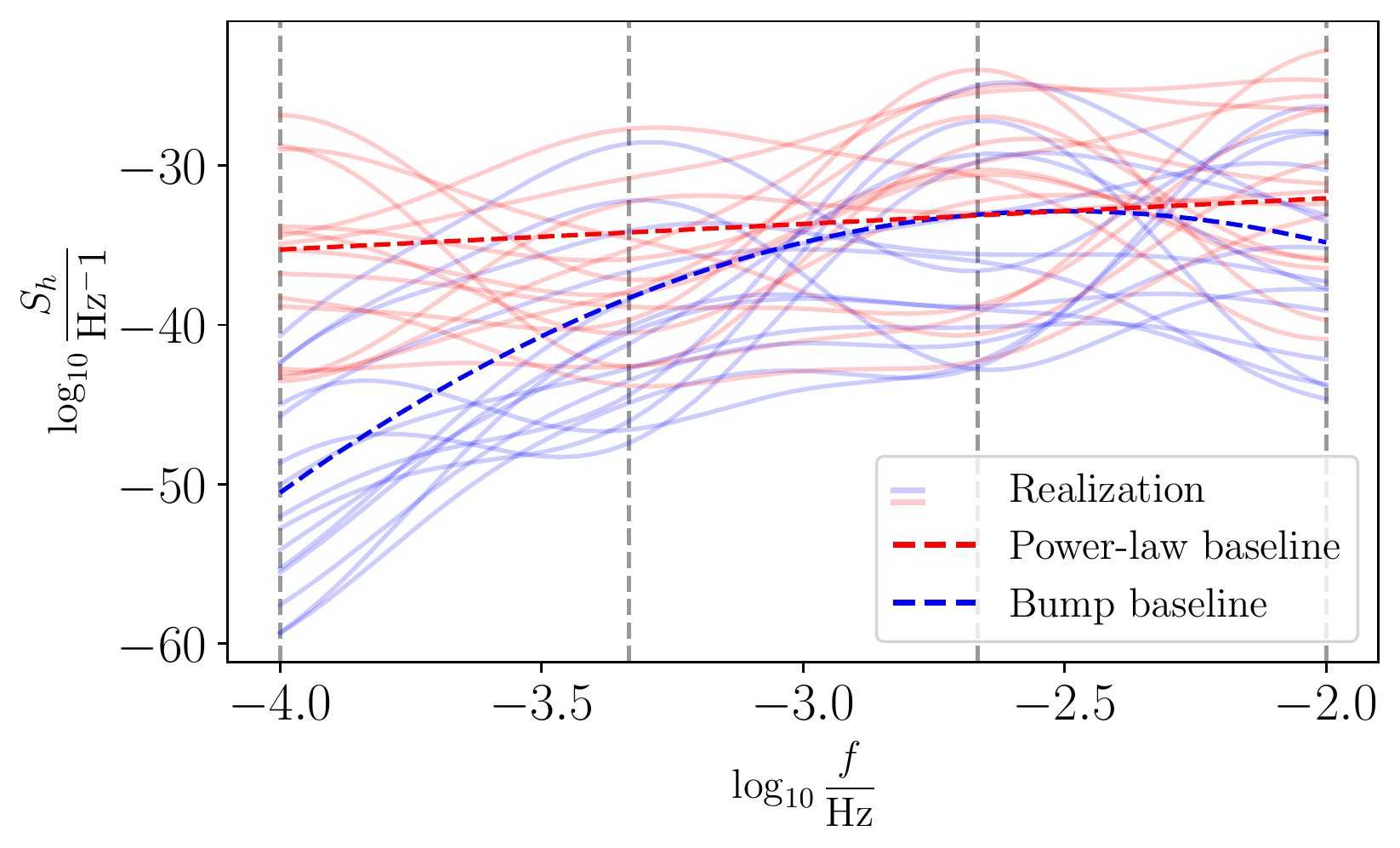}
    \caption{Realizations from two EGP models, with fixed baseline and knots amplitudes sampled from their respective parameter space. Dashed gray lines represent the knot locations. EGP realizations are shown with solid lines, while dashed red (blue) lines correspond to the fixed power-law (bump) baseline.}
    \label{gpreal}
\end{figure}

In this study, we exploit GPs in an unconventional way: we use each GP expectation value from Eq.~\eqref{eq6} as a proposed spectrum, and set $\boldsymbol{g}({X})$ as free parameters. The GP covariance is not used in the inference model.
The dimensionality and domain location of $X$  are to be considered hyperparameters.

Henceforth, we model the logarithm of noise and signal PSDs as two independent EGPs over the logarithmic frequencies, i.e. $X=\left\{ \log_{10} f_1,\dots,\log_{10} f_n\right\}$.
We illustrate both in Secs.~\ref{subsec:signal} and~\ref{subsec:noise}, respectively.
We will refer to the parameters $X$ and to the mean function $m(X)$ as \textit{knots} and \textit{baseline}, respectively.

\subsection{Signal\label{subsec:signal}}
We assume the response matrix $R_{\alpha\beta}(f)$ to be exactly known. 
Thus, we need to specify $S_h(f)$, only.
We use an EGP, with baseline and knots left free to vary simultaneously. 
We do so to be able to capture both global spectral shapes and fine structure in specific frequency regions: 
the choice of baseline influences globally the  spectral-shape proposals, while knots control the fine-local structure.
In Figure~\ref{gpreal}, we show the proposed EGP models for two baseline families, i.e. power-laws and \emph{bump} spectra.
The knots variability can compensate easily for misidentification of the truth baseline family.
In our inference, we use a power-law baseline family for $\log_{10}S_h(f)$, parametrized by logarithmic amplitude $\log_{10} A$ at a reference frequency $f_\star = 10^{-2.5} \rm Hz$ and slope $\gamma$. Thus, $\boldsymbol{\mu}({X}_*)$ in Eq.~\eqref{eq6} reads $\log_{10} A + \gamma \log_{10} (f/f_\star)$. 
Fine-structure deviations with respect to the baseline are parameterized through knots $\boldsymbol \delta^h$, defined by
\begin{align}
    g(X)_k -\mu({X})_k &= \log_{10}(S_h(f_k)\cdot \delta^h_k) - \log_{10}{S_h(f_k)} \nonumber \\
    &= \log_{10} \delta^h_k \label{eq8}
\end{align}
The number of knots is fixed in each inference, and we choose the associated frequencies $f_k$ equally log-spaced within the data frequency range. Overall the model is summarized by functions from the parametric family 
\begin{align}
    &\log_{10} S_h(f;{\log_{10} A, \gamma, \boldsymbol \delta^h}) =  \log_{10}  A \,+ \nonumber\\
    &+\gamma \log_{10} \left(f/f_\star\right) + k(f,f_j)k({f_j,f_k})^{-1}\log_{10} \delta^h_k,
\end{align}

\subsection{Noise\label{subsec:noise}}

\begin{figure}
    \centering
    \includegraphics[width=0.95\columnwidth]{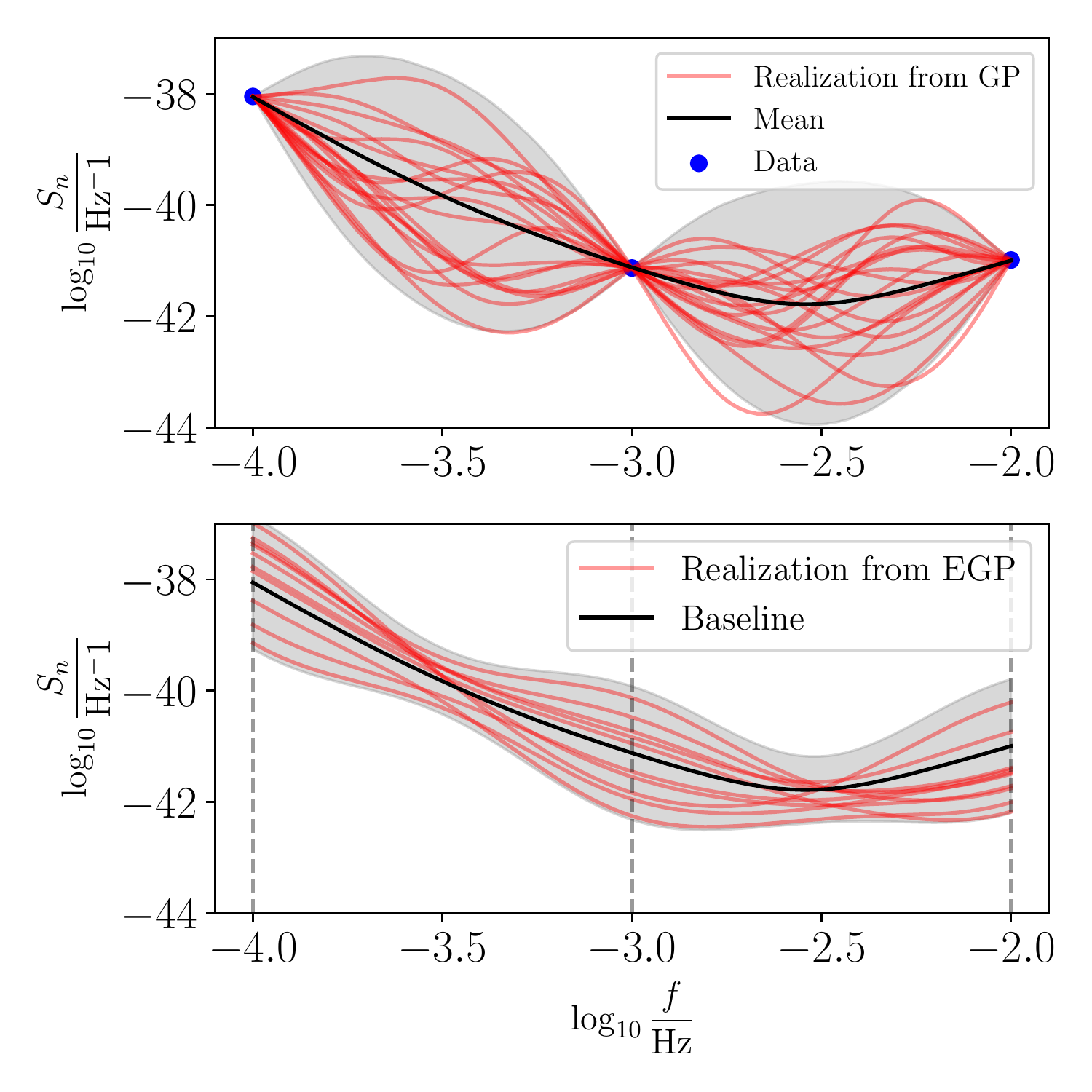}
    \caption{Noise realizations from GP and EGP model. Top panel: realizations from GP. The gray shaded region represents the $90\%$ credible interval. 
    Bottom panel: Realizations from EGP. The gray area represents the prior $90\%$ credible interval.
    While GP allows for straightforward inclusion of observed data, EGP smoothly explores a large set of spectral shapes, with the \emph{data} (i.e. knots amplitudes at chosen locations shown as grey dashed vertical lines) effectively proposed as inference parameters.}
    \label{gpnoise}
\end{figure}

We follow a similar approach to parameterize and infer upon the noise uncertainties. Again, we assume a well known TDI transfer matrix $M_{\rm TDI}$, so the modelling freedom is left for the single-link noise PSD. 
Specifically, we introduce a set of parameter $\log_{10} \boldsymbol {\delta}^n$. 
Contrary to the signal case discussed in Sec.~\ref{subsec:signal}, the baseline is fixed to the nominal PSD level from  ESA’s science requirement document~\cite{LISA2018}.  Thus, the final noise model reads:   
\begin{align}
    \log_{10} S_{n}(f;{{\boldsymbol\delta}^n})&= \log_{10} S^{\rm SciRD}_{n} + \nonumber \\ &+ k(f,f_j)k({f_j,f_k})^{-1}\log_{10} \delta^n_k,
\end{align}

With this parameterization, we augment the noise reference model with considerable flexibility to vary across a large functional space. 
In all our inference the injected noise in the data is generated according to the fixed baseline, so we anticipate recovering zero values for $\log_{10} \boldsymbol \delta^n$. In Fig.\ref{gpnoise}, we show a comparison between GP- and EGP-like instrumental noise modelling, illustrating the different degree of flexibility each one exhibits.  

\subsection{Likelihood}
Inference is performed simultaneously on SGWB and instrumental noise. We construct joint posteriors on parameters $\boldsymbol{\theta}$
\begin{equation}
    p(\boldsymbol{\theta}|\Tilde{d}) \propto \mathcal{L}(\Tilde{d}|\boldsymbol{\theta})\pi(\boldsymbol{\theta})
\end{equation}
through stochastic sampling of the likelihood $\mathcal{L}(\Tilde{d}|\boldsymbol{\theta})$ under chosen priors $\pi(\boldsymbol{\theta})$. 
To do so, we use  \textsc{Balrog}, a large codebase for simulation and inference on LISA signals.
In this study, we use it alongside a nested sampling algorithm as implemented in \textsc{Nessai}~\cite{2021PhRvD.103j3006W} to obtain each posterior and evidence. 
Data are distributed according to the Gaussian likelihood~\cite{Cutler1994}
\begin{align}
&\log\mathcal{L}(\tilde{d}|\boldsymbol{\theta})\! = \\
&  = -\frac{1}{2} \displaystyle \sum_{\alpha,\beta}
\displaystyle \sum_{f=f_{\rm min}}^{f_{\rm max}} \frac{\Tilde{d}_{\alpha}(f)\Tilde{d}^*_{\beta}(f)}{S_{n,\alpha\beta}(f,\boldsymbol \delta^n) + S_{h,\alpha\beta}(f,\log A,\gamma,\boldsymbol \delta^h)}\Delta f \nonumber,
\end{align}
with $\boldsymbol{\theta} = (\log_{10} A,\gamma,\log_{10}\boldsymbol \delta^h,\log_{10}\boldsymbol \delta^n)$.
We choose uniform priors for each parameter in $\boldsymbol \theta$, and the following prior ranges:
\begin{itemize}
    \item Power-law amplitude $\log_{10} A$: $[-70, -35]$
    \item Power-law slope $\gamma$: $[-5, 5]$
    \item Signal knots amplitudes $\log_{10} \boldsymbol{\delta}^h$: $\left[-2, 2\right]$
    \item Noise knots amplitudes $\log_{10} \boldsymbol \delta^n$: $[-0.5, 0.5]$
\end{itemize}
Additionally, we will consider a prior uniform in $\left[-5, 5\right]$ for $\log_{10} \boldsymbol{\delta}^h$, instrumental to the results discussion in Sec.~\ref{sec:result}.

We highlight once again that the number of knots $\boldsymbol \delta^h$ and $\boldsymbol \delta^n$ can be further optimized, as much as the remaining hyperparameters (i.e. the kernel length scale, its functional form, and the knots locations).
This represent a potential high-degree of flexibility to our algorithm, which can be leveraged to find unexpected spectral features upon inference.
In this study, we focus on a finite set of kernel lengths  $\sigma$ and number of knots.
Noise parameter estimation is instead fixed to three knots, given the baseline choice matching the noise model used for injection in the data. 

In absence of a trans-dimensional sampling framework \cite{Green1995} in our codebase, our study is constrained to perform inference for each choice of hyperparameters, and select the most suitable via Bayesian model selection.
In order to compare the different choices, we use the log-Bayes factor
\begin{equation}
    \log_{10}\mathcal{B}^{\sigma,n}_{\sigma^\prime,n^\prime} = \log_{10}\mathcal{Z}_{\sigma,n} - \log_{10}\mathcal{Z}_{\sigma^\prime,n^\prime},
\end{equation}
where $(\sigma,n)$ and $(\sigma^\prime, n^\prime)$ are labels identifying competing models with different kernel length scale and number of knots, respectively, and 
\begin{equation}
    \mathcal{Z} = \displaystyle \int d\boldsymbol{\theta} \mathcal{L}(\Tilde{d}|\boldsymbol{\theta})\pi(\boldsymbol{\theta})
\end{equation}
is the Bayesian evidence.
In Sec.~\ref{sec:result} we present our results, together with a map of evidences: they are to be interpreted as the model marginal likelihoods, conditioned on hyperparameters, only.
This is a core building block towards a trans-dimensional exploration of the full EGP model for SGWB inference and its inclusion in a larger global fit scheme \cite{Karnesis2023,Cornish2015}.

\section{Inference Results}
\label{sec:result}
In this section, we present parameter estimation results performed using the model introduced in Sec.~\ref{sec:stat-model}. First, we use a simple toy model to control the correct recovery of a SGWB with known spectral shape (Sec.~\ref{subsec:toy-model}). Then, we apply our formalism to the parameter estimation of an SGWB from a population of unresolved EMRIs (Sec.~\ref{subsec:emri-sgwb}).
Throughout the analysis, LISA data are simulated for 4 years of observation time. We proceed with the assumption that non-stochastic GWs sources and glitches have been subtracted perfectly. Additionally, we consider the absence of data gaps, resulting in the generation of an idealized residual dataset.

\subsection{Toy Model\label{subsec:toy-model}}

We conduct a test retrieving an EGP model with a power-law baseline and non-zero knot amplitudes. 

\begin{figure}
    \centering  
\includegraphics[width=0.95\columnwidth]{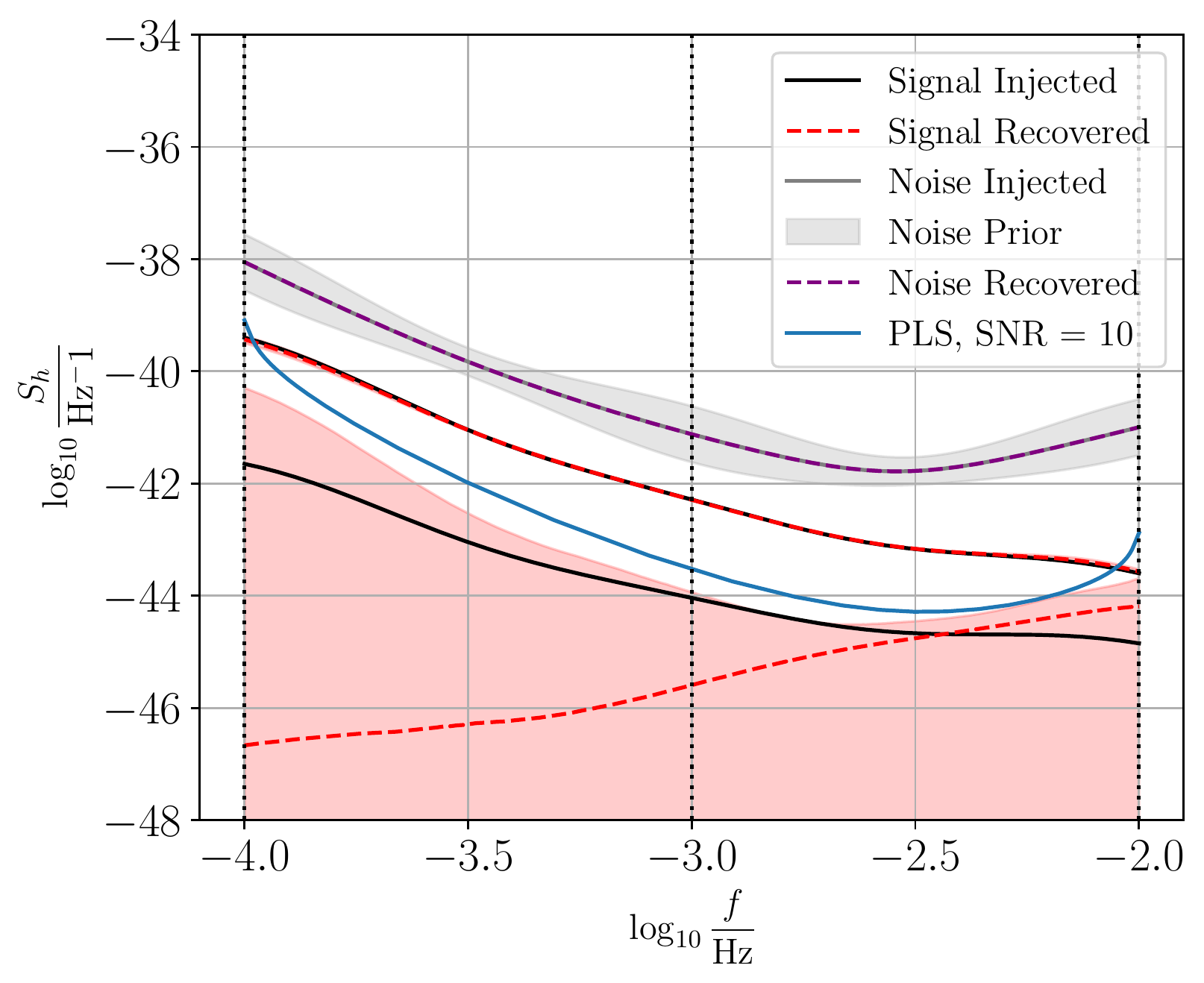}
\caption{
    Spectral reconstruction of two toy model SGWBs.
    Injected SGWBs are denoted by black solid lines. They have SNR of 209 and 5, respectively. This places them above and below the nominal powerlaw sensitivity curve (solid blue line) to an SGWB with SNR of 10 after 4 years of observation. 
    Noise injection and prior 90$\%$ confidence intervals are shown as gray solid and purple dashed lines, respectively.
    Posterior median and $90\%$ confidence intervals on each signal --analysed independently-- are denoted by red dashed lines and light-red shared areas. 
    The EGP flexibility captures features in the PSD shape.}
\label{fig:plottoy}
\end{figure}

Although the injected signals lack an astrophysical interpretation, they serve as a compelling test for our method's capability of recovering complex spectral shapes. 
We inject separately two SGWBs, with SNR of 209 and 5, respectively. We choose the following parameters for each signal:
\begin{itemize}
    \item Power-law baseline with $\log_{10} A=-43, \gamma=-2$ (SNR=209) and $\log_{10} A=-44.5, \gamma=-1.5$ (SNR=5);
    \item Knots amplitudes $\log_{10} \boldsymbol \delta^h = \left[0.6, -0.2, -0.3, 0.4\right]$, identical in both cases.
\end{itemize}
We choose the parameters as representative of two SGWBs, above and below the nominal power-law sensitivity curve~\cite{2013PhRvD..88l4032T} to an SGWB with SNR of 10 after 4 years of observation, respectively.
We emphasize once more that the injected model is an EGP with local features arising with respect to a simple power-law. In this specific context, the PLS is not a faithful indicator of a signal detectability, but should be interpreted as a rough reference level.

Spectral inference results and parameter posteriors are shown in Fig.~\ref{fig:plottoy} and Fig.~\ref{cornertoy} in App.~\ref{Appendix2}, respectively.
The EGP model is able to capture the spectral shape of the high SNR injected signal, and posterior distributions are consistent with the injected values within their $90\%$ credible intervals.
The posterior distributions for the noise parameters are consistent with zero, as expected.
If the signal SNR is too low, as for the second simulated SGWB (SNR=5) the posterior distribution for its parameters is effectively an upper limit, only.

\subsection{Astrophysical Case: EMRI SGWB \label{subsec:emri-sgwb}}
We further apply our method to the astrophysical case of an SGWB from EMRIs. 
We inject a background signal chosen from a set available in literature~\cite{Pozzoli2023}: 
twelve background signals from astrophysically motivated populations~\cite{Babak2017} were generated (therein referred to as $\textsc{M1}$ to $\textsc{M12}$). 
The stochastic signal is constructed through suitable processing of whole population signals, and subtraction of resolvable sources via an iterative algorithm~\cite{Karnesis2021}. Each source GW strain is computed using the state-of-the-art EMRI waveforms~\cite{Katz2021} and the resulting LISA signal is simulated through an accurate response model.
Resulting SGWBs exhibit a wide range of amplitudes, though the majority of them provides an SGWB potentially detectable with LISA over 4 years of observation time.

In this study, we use $\textsc{M1}$ as our fiducial model, as it provides an intermediate SGWB amplitude across the populations studied in~\cite{Pozzoli2023}.
Because of uncertainties on spectral morphology and population distribution, our flexible inference method is suitable to analyse such signal. 
In order to inject coherently the signal in the \textsc{Balrog} simulator, we divide the A channel signal realization obtained by~\cite{Pozzoli2023} by the response function, following the formalism in Eq.~\eqref{eq:signal-spectrum}.
Then, we smooth the resulting spectrum it to obtain a reference model PSD. 
Subsequently, we re-generate each TDI channel data and use it for inference.

\subsubsection{Exploring Hyperparameters Space}

We perform a set of Bayesian inferences, exploring two hyperparameters for the signal model: the number of knots, $n = 3, 4, \ldots, 8$, and the kernel length-scale, as integer $m_\sigma$ multiples of $\sigma = 0.6$, with $m_\sigma = \{1,2,4,8,16,32\}$. 

\begin{figure}
    \includegraphics[width=0.95\columnwidth]{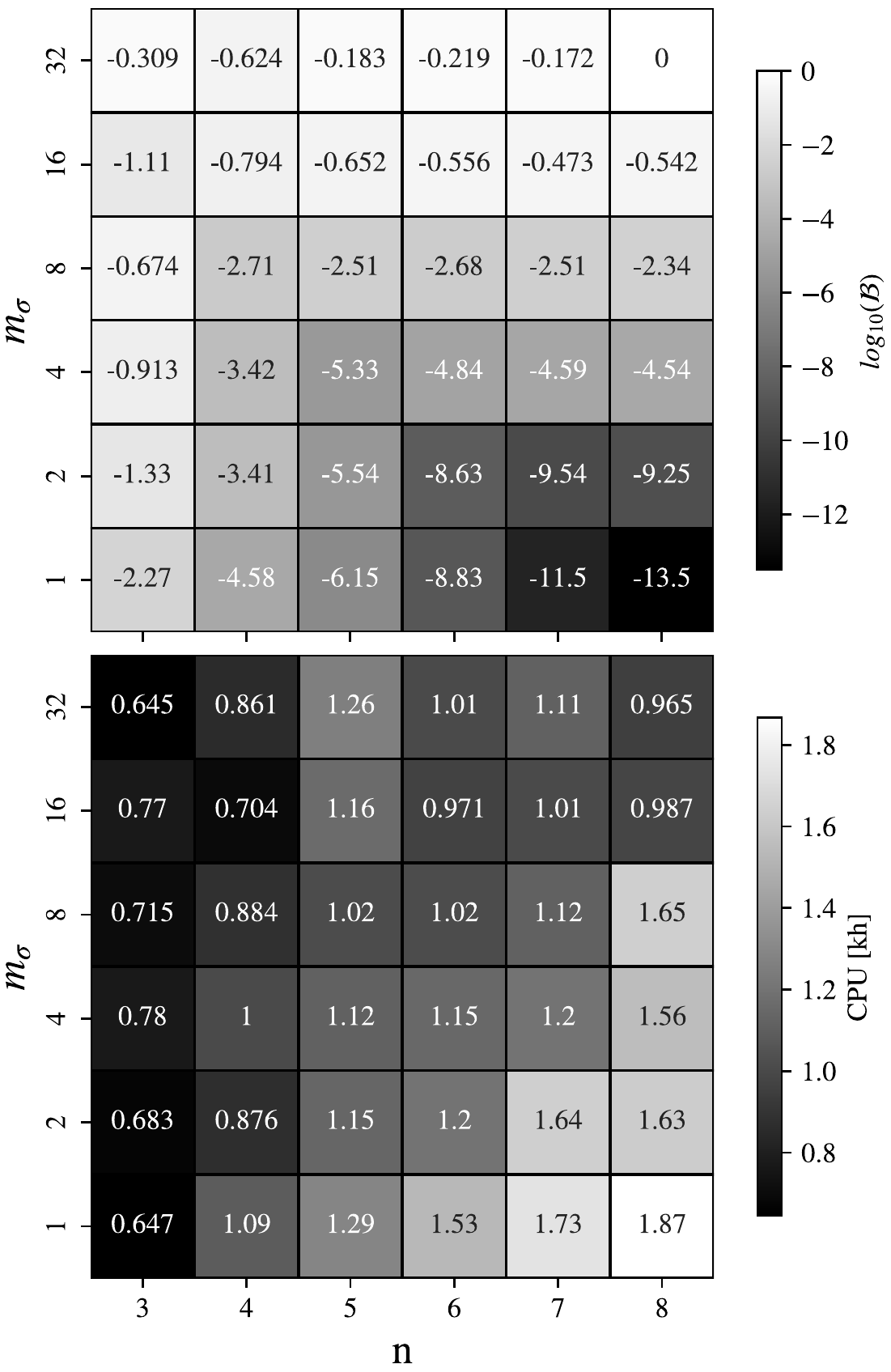}
    \caption{Top Panel: Grid of logarithmic Bayes factor for different pairs of number of signal knots ($n$) and integers of Kernel length scale $\sigma$. The numbers are computed in relation to the model that has the highest evidence, specifically $\left( n, m_\sigma \right) = (8, 32)$. We observe a diagonal pattern: as $m_\sigma$ decreases and $n$ increases, the model becomes progressively less preferred.
    Bottom Panel:  Computational Cost (i.e. Sampling time as reported by \textsc{Nessai}) of parameter estimations as function of integer multiple of $\sigma$, $m_\sigma$ (with $\sigma = 0.6$) and the number of knots of the signal model, $n$. The numbers are in unit of kilohours. Each inference is run in multi-threading over 40 cores.
    }
    \label{bayes}
\end{figure}

\begin{figure}
    \centering
    \includegraphics[width=0.95\columnwidth]{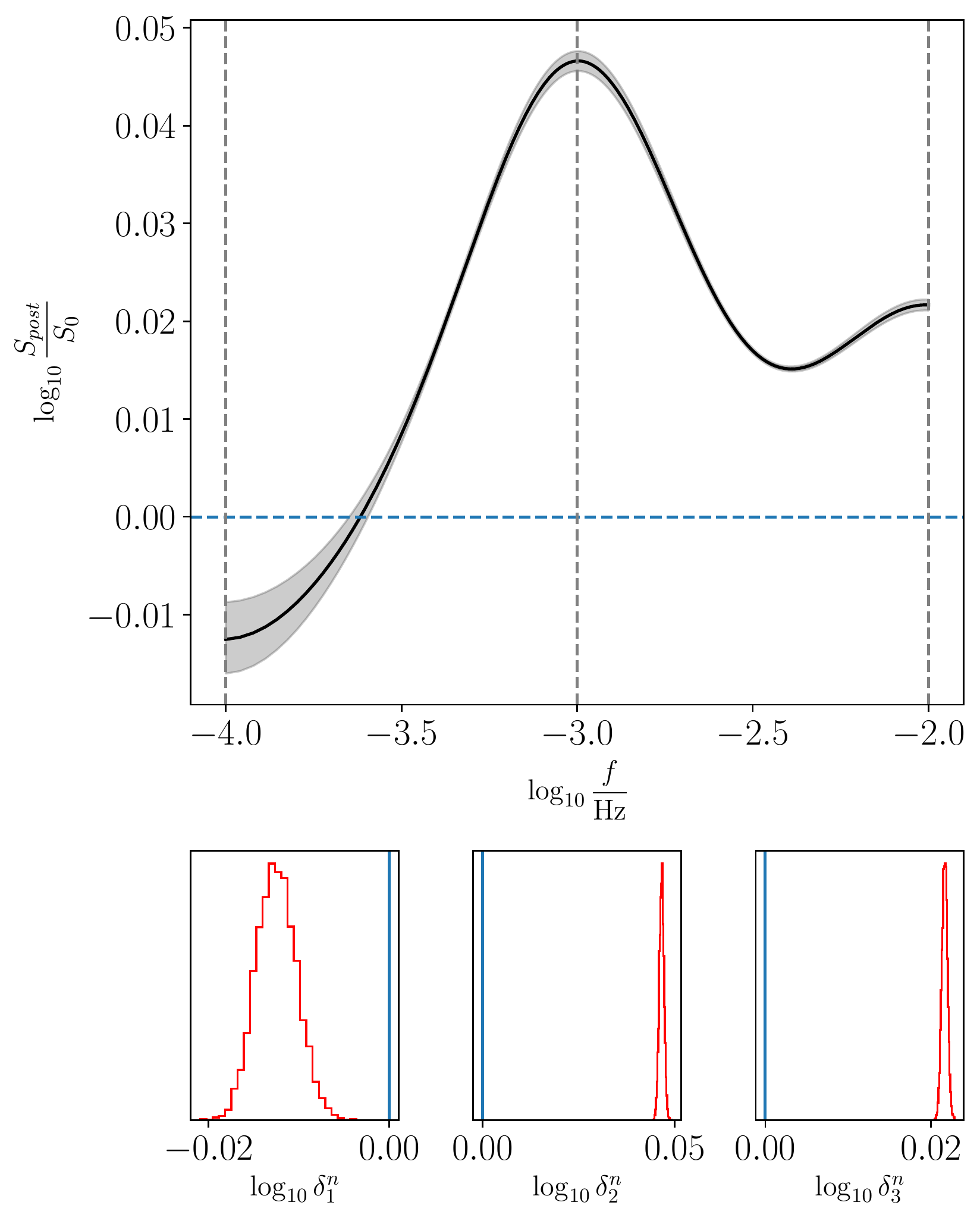}
    \caption{Posterior distribution for the inference on EMRIs background, with a noise-only model. The noise estimation employs 3 knots, with frequencies corresponding to vertical dashed gray lines in the main panel. Bottom sub-panels show posteriors on individual knot amplitudes. The injected spectral noise model corresponds to $\boldsymbol{\delta}^n = \boldsymbol{0}$, and is denoted by solid blue lines. The noise model can compensate with significant biases the presence of a SGWB in the data. Nonetheless, its evidence is significantly lower when compared to the noise and signal models, as illustrated in Fig.~\ref{bayes}. 
    }
    \label{fig:noise2}
\end{figure}

We evaluate each evidence through stochastic nested sampling. Results are presented in Fig.~\ref{bayes} (top panel), as Bayes factors $\log_{10} \mathcal{B}$ with respect to the highest-evidence model, i.e. $\left(n, m_\sigma \right)= (8, 32)$. 
They reveal a prevailing trend:
as $m_\sigma$ increases, the number of knots become less effective at influencing each model evidence, resulting in Bayes factors increasingly close to each as function of $m_\sigma$.

For completeness,in Fig.~\ref{bayes} (bottom panel) we also provide the computational cost of each inference, measured in CPU core kilohours (kh). 
The observed trend is the result of two competing factors: as the number of parameters (i.e. the number of knots $n$) increases, the inference process would take longer.
However, if the increased model complexity (e.g. shorter kernel length-scale) is not required by the simulated data, exploration of parameter space is typically fast. 
As a result, we observe a pattern in computational cost broadly similar to the log Bayes factors: as the number of knots ($\sigma$) increase (decreases), the computational cost diminishes. 

As a further check we perform inference for the null-hypothesis ${\cal H}_0$, where we exclusively model the data as instrumental noise.
Results reveal decisive evidence, in favour of any signal-including model $\left(n, m_\sigma\right)$ with respect to the noise-only model ${\cal H}_0$, with log-Bayes factors largely greater than $10^3$. As expected, the spectral noise reconstruction in Fig.~\ref{fig:noise2} shows significant biases to compensate for the unmodelled SGWB signal in the data.

\subsubsection{Spectral reconstruction}\label{subsec:spectral}

\begin{figure}
    \centering
    \includegraphics[width=0.95\columnwidth]{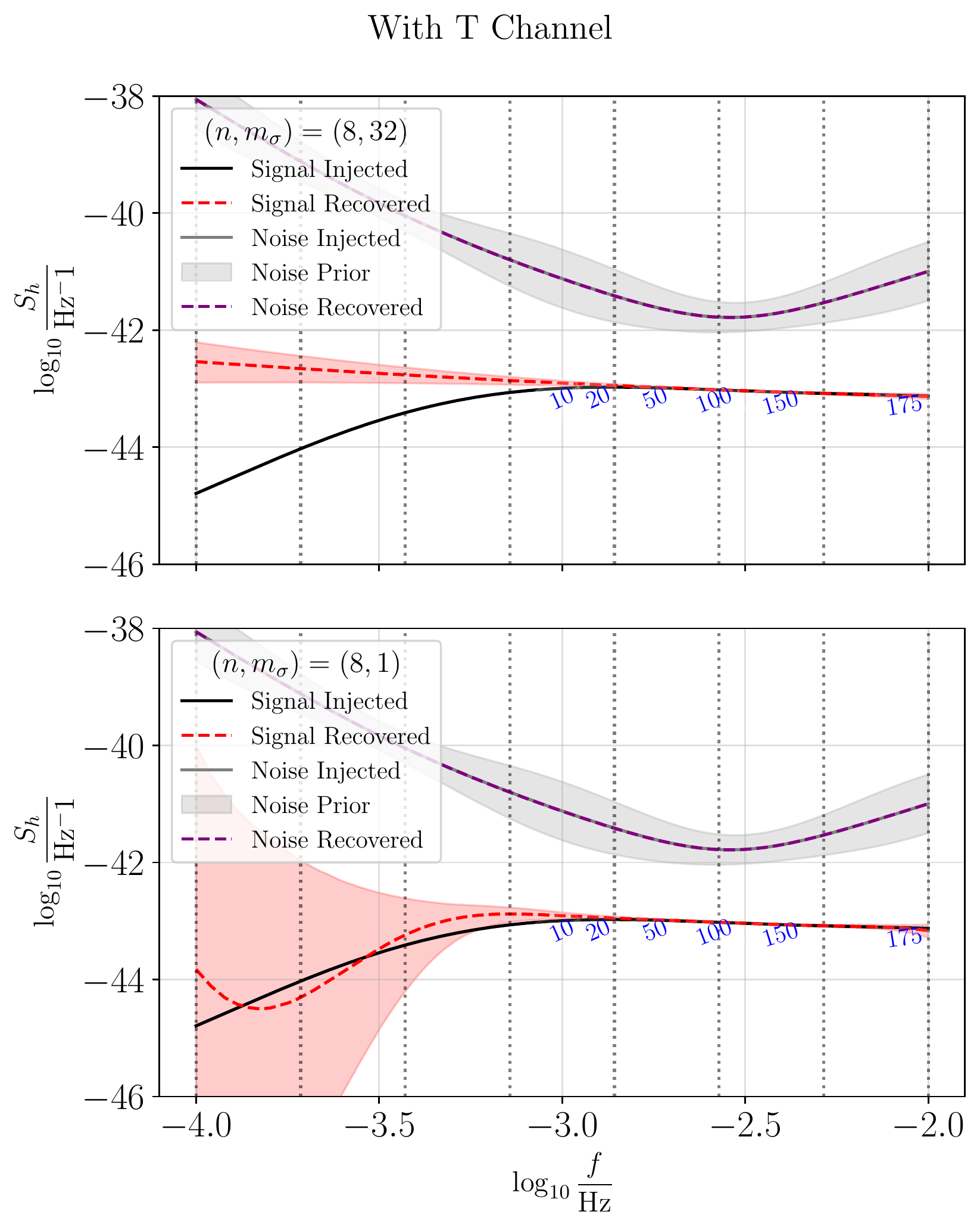}
    \caption{Signal reconstruction  of the M1 EMRI background carried out using the EGP parametrization. In both figures, the noise estimation employs 3 knots. 
    Top Panel: the model features  8 knots for the signal and a kernel length scale that is 32 times larger than $\sigma$. This configuration aligns with the model exhibiting the highest evidence.
    Bottom Panel: the model features 8 knots for the signal, and the kernel length scale is set equal to $\sigma$. This configuration corresponds to the model with the lowest evidence.  Blue numbers denote the cumulative SNR integrated from $10^{-4}\rm{Hz}$ up to the frequency they are located at.}
    \label{fig:emri}
\end{figure}

We discuss here in greater detail the signal spectral reconstruction of each model. 
Our findings are best described by a comparison between the inferences with $(n, m_\sigma)$ equal to $(8,1)$ and $(8,32)$, whose results are shown in top and bottom panels of both Figs.~\ref{fig:emri} and~\ref{fig:emri2}, respectively.
We show the reconstruction of the fiducial EMRI background with and without T channel data in Figs.~\ref{fig:emri} and~\ref{fig:emri2}, respectively. 
The T channel is known to be less sensitive to GWs than the A and E channel at the frequencies of our analyses, therefore acting as a potential noise estimator. 
We observe however that the spectral reconstructions with and without the T channel are broadly consistent, with broader posteriors when the T channel is excluded, as expected.
This is due to the common noise modelling of the single link PSD spectrum, which is then propagated coherently to each channel through $M_{\rm TDI}$ Eq.~\eqref{eq:noise-spectrum}. 
Relaxing the exact knowledge of the transfer matrix is expected to significantly broaden the noise PSD posterior.

A comparison between top and bottom panels reveals an important result, originating from the competition in Bayes factors between excess model complexity and the effectiveness of our chosen parameterization. 
The inference presented in this study is inherently integrated across a large frequency band: we annotate the injected SGWB with blue values for cumulative SNRs with respect to the true noise level on a set of representative frequencies.
It is readily apparent that $80\%$ of the $\rm SNR$ is accumulated between $1$ and $4 {\rm mHz}$: in this frequency range the reconstruction is consistent with the injected spectrum. The dashed red lines and red shaded area denote the posterior median and $90\%$ confidence interval respectively.

At the lowest frequencies, the preferred model (top panel, $n,m_\sigma=8,32$) exhibits larger biases as compared to the disfavoured one (bottom panel, $n,m_\sigma=8,1$).
This is due to the narrow-band ($m_\sigma=1$) fluctuations being highly disfavoured by the data in the high SNR region, where the spectrum is close to a powerlaw model, a shape already described by the baseline parameters.
On the contrary broad-band ($m_\sigma=32$) fluctuations can be absorbed by an adjustment of the baseline parameters, while keeping an accurate spectral reconstruction.
However, the preference for $m_\sigma=32$ has an important side effect: $m_\sigma$ is fixed and equal for every knot in each analysis, including the four at the lowest frequencies. 
Therefore, the model with the highest evidence is not capable of introducing local fluctuations to capture significant tilts with respect to the baseline powerlaw. These are instead provided by the disfavoured model.

Figs.~\ref{fig:corner2} and ~\ref{fig:corner3} in Appendix further support our interpretation, where we compare the posterior distributions for the two runs shown in Fig~\ref{fig:emri} (red contours) with two more differing only by a broader prior for the knot amplitudes $\boldsymbol\delta^h$, uniform in $\left[-5,5\right]$ (blue contours).
The inference results for the noise are largely unaffected and uncorrelated to the signal parameters (and hyperparameters).
However, we find that relaxing the knot priors affects the posterior on the baseline parameters: a sign of strong correlation. 
Moreover, posteriors on $\boldsymbol\delta^h$ for the preferred runs (Fig.~\ref{fig:corner2}) are prior dominated, while the likelihood becomes informative in the disfavoured models $(n,m_\sigma)=(8,1)$ (Fig.~\ref{fig:corner3}).     
In conclusion we argue that in a realistic scenario models more flexible than EGP should be developed, alongside astrophysically motivated priors to avoid introducing --even in an evidence based model selection-- unwantend biases.

\subsubsection{T Channel Influence}
\begin{figure}[ht]
    \centering
    \includegraphics[width=0.95\columnwidth]{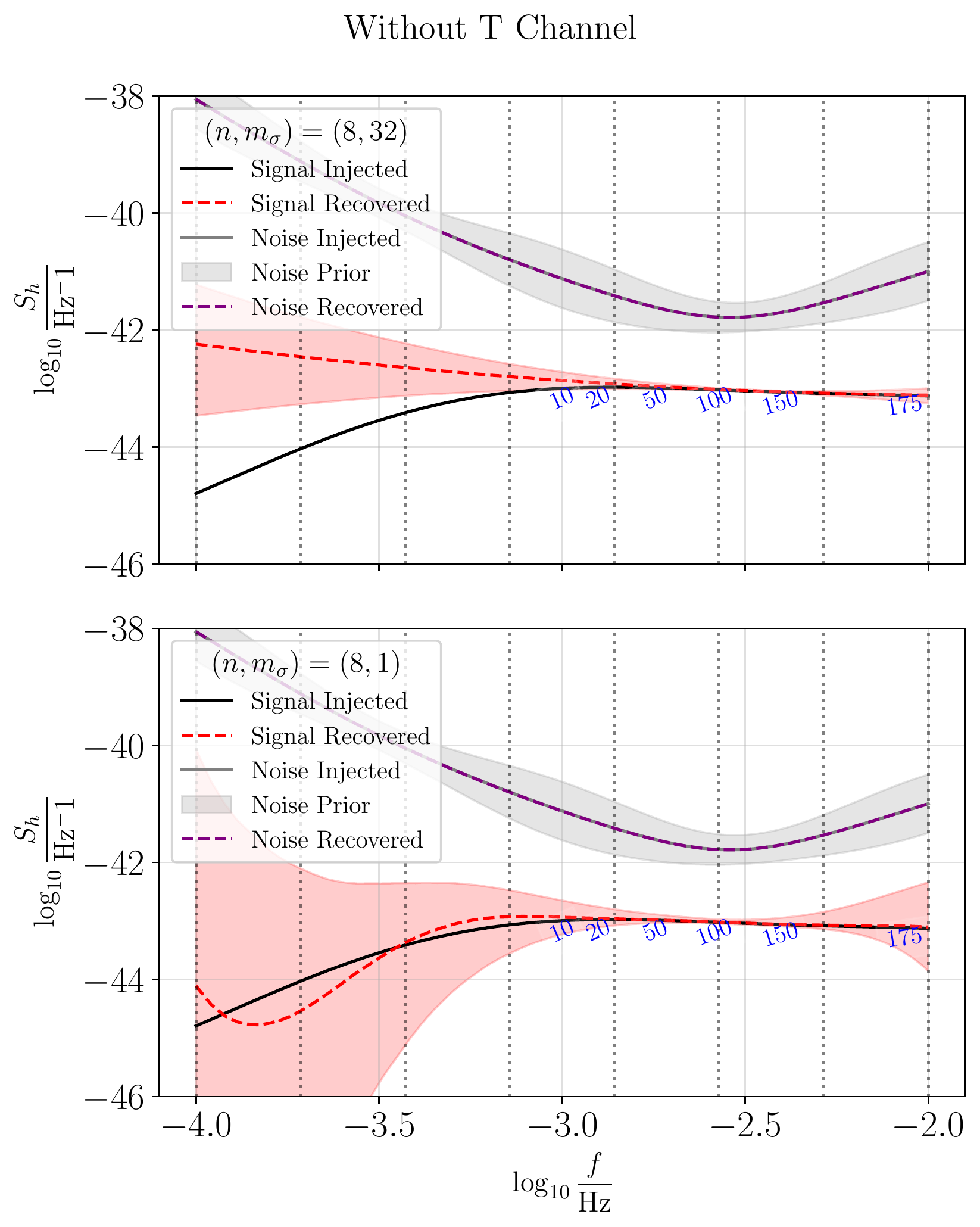}
    \caption{Analogue illustration of Fig. \ref{fig:emri} where the inference has been conducted without incorporating the T channel in the data. Consequently, the resulting posterior regions exhibit greater dispersion, signifying reduced informativeness. Blue numbers denote the cumulative SGWB SNR integrated from $10^{-4}\rm{Hz}$ up to the frequency they are located at.}
    \label{fig:emri2}
\end{figure}

We aim to investigate the impact of the T channel on parameter estimation. Given its reduced sensitivity to the SGWB, we anticipate that excluding the T channel from data analysis will result in less informative posterior distributions (as depicted in Fig. \ref{fig:emri2}). To be more specific, we seek to assess whether hyperparameterization modeling remains influential, as depicted in Fig. \ref{bayes}. 

To achieve this, we examine two models characterized by the highest and lowest values of evidence (or hyperlikelihood), denoted as $\left(n, m_\sigma\right) = \left(8, 32\right)$ and $\left(n, m_\sigma\right) = \left(8, 1\right)$, respectively. 
The calculated logarithmic Bayes factor stands at $-10.95$, underscoring a substantial strength of evidence in favor of one model over the other. 
In other words, the outcome suggests that the trend observed in Fig. \ref{bayes} is robust and not significantly affected by variations in our ability to estimate the noise level. 
\section{Conclusion}
\label{sec:conclusion}
In this study we present a novel method to estimate parameters of an SGWB observed by LISA. 
We implement into our analysis in \textsc{Balrog}, a broader data analysis framework for LISA.
This is a building block towards the completion of a full global fit of deterministic and stochastic GW signals, alongside a flexible noise model. 
We employ a versatile parameterization to infer the SGWB of uncertain spectral shape using EGPs. 
Our method is able to capture unexpected  features in the signal shape, while assessing their significativity in a purely Bayesian formalism. 
We test our algorithm using a simplified toy-model, where the true injected model is by construction within the parameter space explored. 

Following, we apply the analysis to the expected signal from a population of unresolved EMRIs. 
We leverage the model flexibility to (i)explore the hyperparameter space and optimize them based on evidence. 
We also investigate the impact of T channel on the inference results.
We find that the EGP model has great flexibility at reconstructing the SGWB spectral shape. However, while Bayesian evidence might be suitable to select models in the high SNR region, it may disfavour less biased models in the low SNR region, at the price of excess complexity where it is not needed. 
Therefore, the model inherent non-locality should be treated with caution when interpreting spectral results.
We also find that the T channel has a significant impact on the inference results, and its exclusion leads to naturally less constrained results.

Our findings suggest that parametric or weakly-parametric models (as the EGP) should be considered a first step towards even more agnostic and realistic approaches. Including cross-correlation terms between TDI channels is expected to provide a more robust inference, and it is a natural extension of our work.
Finally, for specific SGWBs a parameter estimation modelling non-stationarities and non-Gaussianities would extract valuable information from the data, thus further reducing biases.

\begin{acknowledgments}
We wish to thank Jonathan Gair, Antoine Klein, Golam Shaifullah, Germano Nardini, Chiara Caprini, Mauro Pieroni, Nikos Karnesis, Fabio Rigamonti and all \textsc{Balrog} developers for fruitful discussions. We also wish to thank the University of Geneva for the kind hospitality in summer 2023.
R.B. acknowledges the support of the ICSC National Research Center funded by NextGenerationEU, and the Italian Space Agency grant \emph{Phase A activity for LISA mission}, Agreement n.2017-29-H.0.
C.J.M. acknowledges the support of the UK Space Agency grant, no.\ ST/V002813/1.
A.S. acknowledges the financial support provided under the European Union’s H2020 ERC Consolidator Grant ``Binary Massive Black Hole Astrophysics'' (B Massive, Grant Agreement: 818691).
Computational resources were provided by University of Birmingham BlueBEAR High Performance Computing facility, by CINECA through EuroHPC Benchmark access call grant ``LISAFIT'' (proposal ID: EHPC-BEN-2023B08-024), and by the Google Cloud Research Credits program with the award GCP19980904.

\textit{Software}:
We acknowledge usage of 
\textsc{Mathematica}~\cite{Mathematica} 
and of the following 
\textsc{Python}~\cite{10.5555/1593511} 
packages for modeling, analysis, post-processing, and production of results throughout:
\textsc{Nessai}~\cite{2021PhRvD.103j3006W},
\textsc{matplotlib}~\cite{2007CSE.....9...90H},
\textsc{numpy}~\cite{2020Natur.585..357H},
\textsc{scipy}~\cite{2020NatMe..17..261V}.

\end{acknowledgments}
\clearpage

\appendix
\onecolumngrid

\section{SGWB response function}\label{Appendix}
For completeness and consistency of definitions, we provide here a derivation of the one-arm LISA response --induced by a GW propagating along $\hat{k}$-- as experienced along a single link identified by the sender and receiver spacecraft $s$ and $r$ and the unit-vector pointing from the former to the latter, i.e. $\hat{r}_{sr}$. Henceforth, no summation over repeated $s$ and $r$ is assumed. Each fractional armlength variation reads
\begin{equation}\label{eq:deltaloverl}
    W_{sr}(t,\hat{k})= \frac{\delta L_{sr}(t)}{L_{sr}} \\
    = \frac{1}{2} \hat{r}^a_{sr} \hat{r}^b_{sr} \int^\infty _{-\infty}\Tilde{h}_{ab}(f)\tau_{sr}(f,t,\hat{k})e^{2\pi i f(t- \Delta t)}df,    
\end{equation}
where 
\begin{equation}
    \Delta t = \frac{1}{2}\left[\hat{k}\cdot (\vec{x}_s + \vec{x}_r) + L_{sr} \right]
\end{equation}
and
\begin{equation}
    \tau_{sr} (f,t,\hat{k}) = {\rm sinc} (\pi L_{sr} f [1-\hat{k}\cdot \hat{r}_{sr}(t)]).
\end{equation}
where $\vec{x}_s$ and $\vec{x}_r$ denote the sender and receiver spacecraft, respectively.
Due to the motion of the satellites,
the arm direction vector $\hat{r}$ varies with time.
The GW strain can be written in TT gauge as:
\begin{equation}
    h_{ab}(t,\hat{k}) = h_+(t)\epsilon^+_{ab}(\hat{k}) + h_\times(t)\epsilon^\times_{ab}(\hat{k}).
\end{equation}
where $\epsilon^{+,\times}$ denote a linear polarization-tensor basis.
The stochastic strain is the linear superposition of plane tensor-waves from each direction of the sky, therefore we define an integrated response to each polarization $p =
+, \times$, as follows
\begin{equation}\label{eq:Wintegp}
    W_{sr}^p(t)=\int_{\hat{k}} W_{sr}^p(t,\hat{k}) d^2\hat{k},
\end{equation}
where 
\begin{equation}
\begin{split}
           W_{sr}^p(t,\hat{k}) &=  \frac{1}{2} \hat{r}^a_{sr} \hat{r}^b_{sr} \int^\infty _{-\infty}\Tilde{h}_p(f,\hat{k})\epsilon_{ab}^p(\hat{k})\tau(f,t,\hat{k})e^{2\pi i f(t-\boldsymbol \delta t)}df = \\
        &= \frac{1}{2} \int^\infty _{-\infty}\Tilde{h}_p(f,\hat{k})\xi^p(\hat{k},\hat{r}_{sr})\tau(f,t,\hat{k})e^{2\pi i f(t-\boldsymbol \delta t)}df.
\end{split}
\label{app1}
\end{equation}
In Eq.~\eqref{eq:Wintegp} we introduced the antenna pattern functions $\xi^p(\hat{k},\hat{r}_{sr})$.
They read as:
\begin{equation}
    \begin{split}
        &\xi^+(\hat{k},\hat{r}_{sr}) = (\hat{u}_k \cdot \hat{r}_{sr})^2 - (\hat{v}_k \cdot \hat{r}_{sr})^2 \\
        &\xi^\times(\hat{k},\hat{r}_{sr}) = 2(\hat{u}_k \cdot \hat{r}_{sr}) (\hat{v}_k \cdot \hat{r}_{sr}).
    \end{split}
\end{equation}
We rewrite Eq.~\eqref{app1} equivalently as:
\begin{equation}
    W_{sr}^p(t,\hat{k}) = \int^\infty _{-\infty}\Tilde{h}_p(f,\hat{k}) \overline{G}^{p}_{sr}(f,t,\hat{k})e^{2\pi ift}df,
    \label{app2}
\end{equation}
where $\overline{G}^{p}_{sr}(f,t,\hat{k})$ reads
\begin{equation}
    \overline{G}^{p}_{sr}(f,t,\hat{k}) = \frac{1}{2}\xi^p(\hat{k},\hat{r}_{sr})e^{-i\pi f(L_{sr} + \hat{k}\cdot(x_s + x_r))}\tau_{sr}(f,t,\hat{k}).\label{appGbar}
\end{equation}
Noteworthy, $\overline{G}^{p}_{sr}(f,t,\hat{k})$ is a complex quantity unlike $\tau(f,t,\hat{k})$. 
Equivalently, we recast Eq.~\eqref{appGbar} as a response in fractional frequency domain
\begin{equation}
    G^{p}_{sr}(f,t,\hat{k}) = 2\pi i L_{sr}f \overline{G}^{p}_{sr}(f,t,\hat{k}).
\end{equation}

The Fourier transform of~\eqref{app2} is then
\begin{align}
    W_{sr}^p(f,\hat{k}) &= \int^{\infty}_{-\infty} dt e^{-2\pi i ft} W_{sr}^p(t,\hat{k})  \\
    &= \int^{\infty}_{-\infty} dt e^{-2\pi i ft}  \int^{\infty}_{-\infty} \Tilde{h}_p(f',\hat{k}) \overline{G}^{p}_{sr}(f',t,\hat{k})e^{2\pi if't}df' \nonumber\\
    &= \int^{\infty}_{-\infty} dt \int^{\infty}_{-\infty} \Tilde{h}_p(f',\hat{k}) \overline{G}^{p}_{sr}(f',t,\hat{k})e^{-2\pi it(f-f')}df' \nonumber\\
    &= \int^{\infty}_{-\infty} df' \overline{G}^{p}_{sr}(f',f-f',\hat{k})\Tilde{h}_p(f',\hat{k}) \nonumber,
\end{align}
note that
\begin{equation}
    \overline{G}^{p}_{sr}(f',f,\hat{k}) = \int ^\infty _{-\infty}e^{-2\pi i ft}\overline{G}^p_{sr}(f',t,\hat{k})dt.
\end{equation}
As described in Eq.\eqref{eq:strainpsd} an isotropic, stationary, zero-mean background is characterized entirely by the strain covariance
which allows for the definition of a response matrix for two given links $sr$ and $s^\prime r^\prime$ reads
\begin{equation}    
     \Sigma _{sr,s'r'}(f) = \left< W^p_{sr}(f) W^{*p}_{s'r'}(f)\right> = 
     \int_{\hat{k}}d^2k\int^\infty_{-\infty}df'\overline{G}^p_{sr}(f',f-f',\hat{k})\overline{G}^{*p}_{s'r'}(f',f-f',\hat{k})S_h(f')  
\label{app3}
\end{equation}

For an isotropic distribution of sources, integrating the product of $\overline{G}$'s and evaluating it at a reference time $t_0$ --without loss of generality given the assumption of perfect stationarity-- we obtain
Thus,
\begin{equation}
    \Sigma ^p_{sr,s'r'}(f) =  S_h(f) G^p_{sr,s'r'}(f),
\end{equation}
with
\begin{equation}
    G^p_{sr,s'r'}(f) = \int_{\hat{k}}d^2k\overline{G}^p_{sr}(f,t_0,\hat{k})\overline{G}^{*p}_{s'r'}(f,t_0,\hat{k})
\label{app5}
\end{equation}
which corresponds to $G_{ij}$ in Eq.~\eqref{eq:single_link_response_matrix} upon remapping of the indices $sr,s^\prime r^\prime$ to $ij$.
\vfill
\clearpage

\section{Posterior distribution for the toy model\label{Appendix2}}
\vfill
\onecolumngrid
\begin{figure}[ht]
    \centering
    \includegraphics[width=1.0\columnwidth]{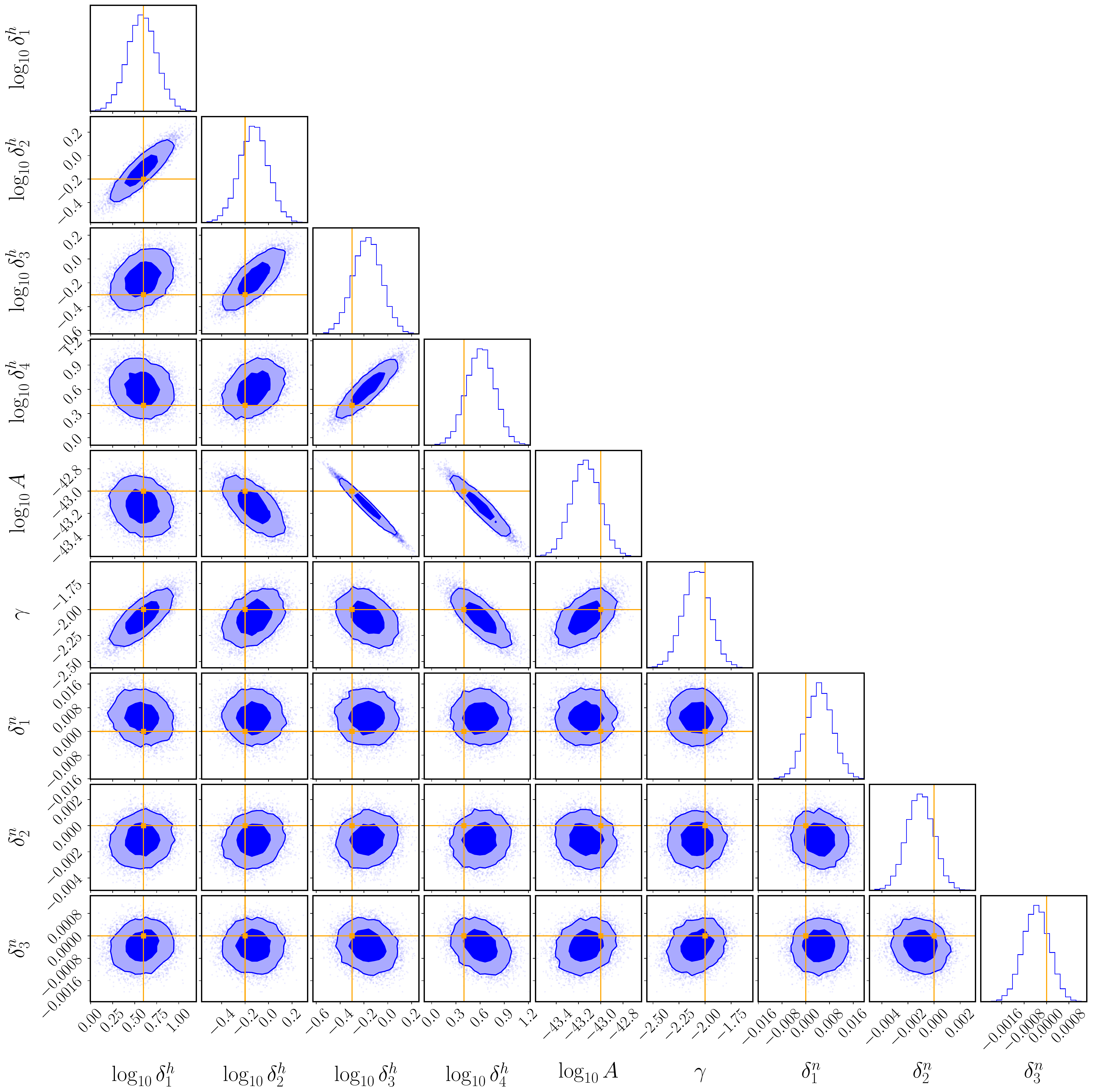}
    \caption{Posterior parameter distribution for the toy-model SGWB with SNR=209 described in Sec.~\ref{subsec:toy-model}, obtained with an EGP model inference. 
    The signal injected parameters are $\left\{\log_{10} \boldsymbol \delta^h, \log_{10} A, \gamma \right\} = \left\{\left[0.6,-0.2,-0.3,0.4 \right],-43,-2 \right\}$. 
    The injected noise level matches the nominal LISA sensitivity curve~\cite{LISA2018}. 
    Darker (lighter) shaded areas denote 90\% (50\%) credible regions, and solid orange lines indicate the injected values.
    Posteriors are consistent with the injected values at the 90\% credible level, and the noise parameters are consistent with zero with accuracies between $10^{-3}$ and $10^{-2}$.
    \label{cornertoy}}
\end{figure}

\vfill
\clearpage

\section{Posterior distributions for Model M1 SGWB}
\vfill
\onecolumngrid
\begin{figure}[h!]
    \centering
    \includegraphics[width=1.0\columnwidth]{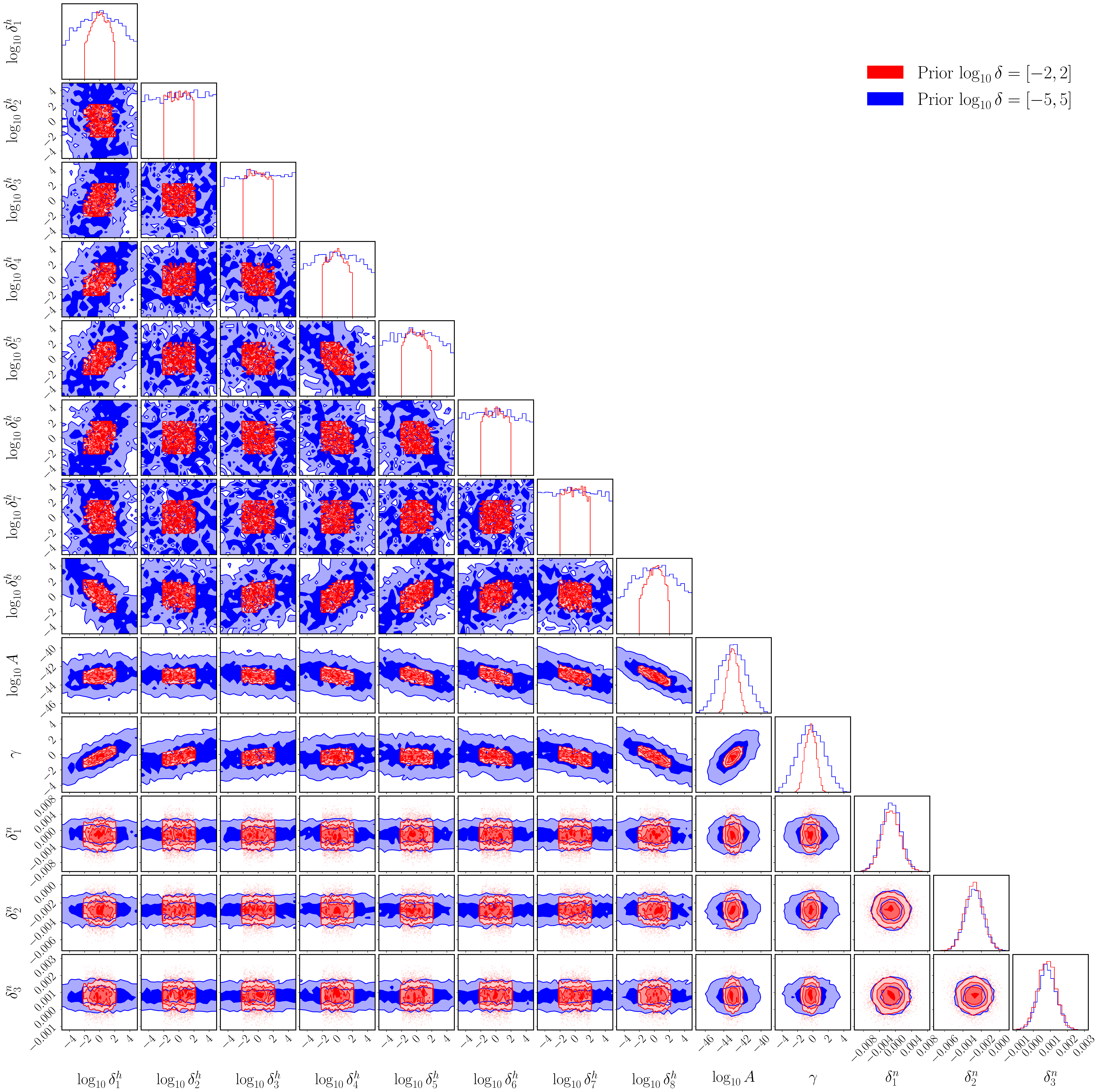}
    \caption{Posterior parameter distribution for M1 EMRI SGWB for model $(n,m_\sigma)=(8,32)$ using a larger and narrower prior on $\boldsymbol{\delta}^h$. Posteriors shown in this plot correspond to the spectral reconstructions in top panels of Fig.~\ref{fig:emri} and ~\ref{fig:emri2},respectively. They support the result interpretation provided in Sec.~\ref{subsec:spectral}.
    \label{fig:corner2}
    }
\end{figure}

\vfill

\clearpage

\vfill

\begin{figure}[h!]
    \centering
    \includegraphics[width=1.0\columnwidth]{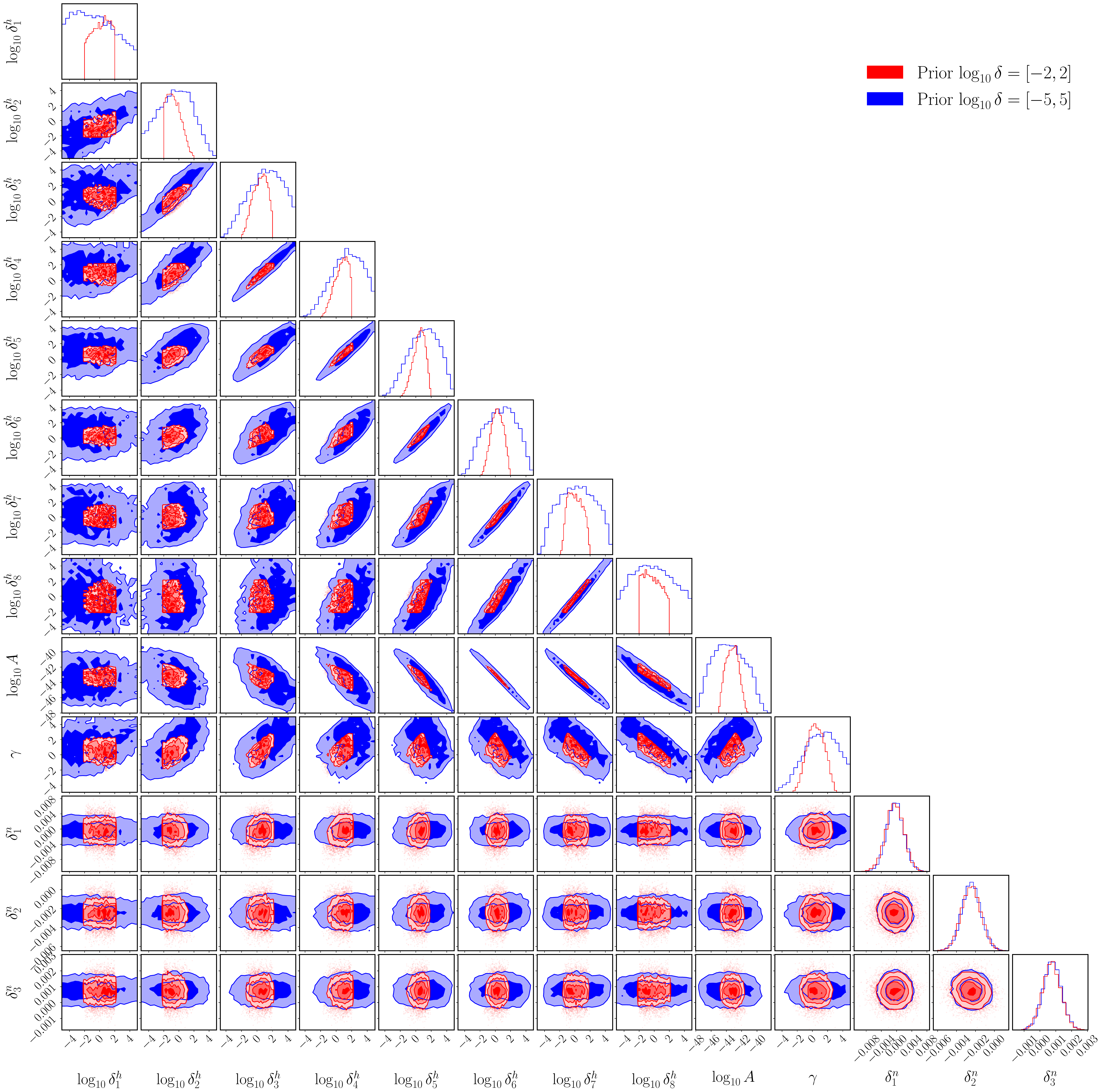}
    \caption{Posterior parameter distribution for M1 EMRI SGWB for model $(n,m_\sigma)=(8,1)$ using a larger and narrower prior on $\boldsymbol{\delta}^h$. Posteriors shown in this plot correspond to the spectral reconstructions in bottom panels of Fig.~\ref{fig:emri} and ~\ref{fig:emri2}, respectively. They support the result interpretation provided in Sec.~\ref{subsec:spectral}.
    \label{fig:corner3}
    }
\end{figure}

\clearpage
\twocolumngrid

\bibliography{biblio}
\bibliographystyle{apsrev4-2}
\end{document}